\documentclass[12pt]{article}
\input epsf.sty
\usepackage{mathrsfs}
\usepackage{amsmath}
\usepackage{mathrsfs}
\usepackage{amssymb}
\usepackage{color}
\usepackage{psfrag}
\newcommand{\bea}{\begin{eqnarray}}
\newcommand{\eea}{\end{eqnarray}}
\newcommand{\be}{\begin{equation}}
\newcommand{\ee}{\end{equation}}
\newcommand{\vs}[1]{\vspace{#1 mm}}

\newcommand{\dsl}{\pa \kern-0.5em /}

\newcommand{\pa}{\partial}

\newcommand{\nn}{\nonumber\\}

\newcommand{\eqn}[1]{(\ref{#1})}

\begin{document}
\topmargin 0mm
\oddsidemargin 0mm

\begin{flushright}

USTC-ICTS-11-05\\

\end{flushright}

\vspace{2mm}

\begin{center}

{\Large \bf The enriched phase structure
 of black branes \\in
canonical ensemble}

\vs{10}

{\large J. X. Lu$^a$\footnote{E-mail: jxlu@ustc.edu.cn}, Shibaji
Roy$^b$\footnote{E-mail: shibaji.roy@saha.ac.in} and Zhiguang
Xiao$^a$\footnote{E-mail: xiaozg@ustc.edu.cn}}

\vspace{4mm}

{\em

 $^a$ Interdisciplinary Center for Theoretical Study\\
 University of Science and Technology of China, Hefei, Anhui
 230026, China\\

\vs{4}

 $^b$ Saha Institute of Nuclear Physics,
 1/AF Bidhannagar, Calcutta-700 064, India\\}

\end{center}

\vs{10}

\begin{abstract}

It is found that a necessary completion of phase structure of
$D$-dimensional charged black $p$-brane ($p
> 0$) in a cavity requires two additional thermodynamical
phases, the so-called ``bubble of nothing'' and/or the extremal
brane, in canonical ensemble. This finding resolves the puzzle about
the missing phases which are needed for the underlying phase diagram
when $\tilde d = D - p -3  \le 2$ and gives a new (bubble) phase
which can become globally stable when $\tilde d > 2$. An analog of
Hawking-Page transition is also found among other new phase
transitions, giving a complete phase structure in this setup.

\end{abstract}
\newpage
\section{Introduction}
Understanding the nature of black hole thermodynamics may teach us
lessons about quantum gravity. The underlying phase structure can be
useful not only in this regard but for other purposes as well. For
example, with the advent of AdS/CFT correspondence, the known phase
structure of large black holes in asymptotically Anti-de Sitter
(AdS) space can be used to understand various physical phenomena in
other branches of physics. An illustration of this is that the
Hawking-Page transition for AdS black hole  `evaporating' into
regular ``hot empty AdS space"  at certain temperature
\cite{Hawking:1982dh} can enhance understanding of the
confinement-deconfinement phase  transition in large $N$ gauge
theory \cite{Witten:1998zw}.

A large part of the phase structure of an AdS black hole
\cite{Chamblin:1999tk, Chamblin:1999hg}  is actually not unique to
the black hole in asymptotically AdS space but shared universally by
suitably stabilized black holes/branes, say, in asymptotically flat
space \cite{Carlip:2003ne, Lundgren:2006kt, Lu:2010xt, Lu:2010au},
even in the presence of a charge $q$. For example, a chargeless
(suitably stabilized) asymptotically flat black $p$-brane can also
undergo a Hawking-Page transition at certain temperature, now
evaporating into a regular `hot flat space' instead. When $q \neq
0$, there exists also a critical charge $q_c$ and for $q < q_c$, the
phase diagram universally contains a van der Waals-Maxwell
liquid-gas type phase structure along with a first-order phase
transition line ending at a second-order critical point with a
universal exponent for the specific heat as $- 2/3$ when $q = q_c$.
This universal phase structure may hint holography even in
asymptotically flat space, as pointed out in \cite{Carlip:2003ne}.

However, unlike an AdS black hole, an isolated asymptotically flat
black hole/brane is unstable due to its Hawking radiation and needs
to be stabilized first before one can discuss the equilibrium
thermodynamics. To establish its stability, the standard practice is
to place such a system inside a finite spherical cavity
\cite{York:1986it} with its surface temperature fixed. In other
words, a thermodynamical ensemble is considered which can
be either canonical or grand canonical, depending on whether the
charge inside the cavity or the potential at the surface of the
cavity is fixed \cite{Braden:1990hw}. In this paper, our focus is
the canonical ensemble, i.e., the charge inside the cavity is fixed,
and in particular our main interest is to study the phase structure
when the flux/charge inside the cavity is fixed but non-zero.

When the ensemble temperature drops below a certain minimum value,
there appears a puzzle of missing phases, in certain cases, as it is
not clear where the system would be in the absence of globally
stable phases (No such issues arise in grand canonical ensemble,
however, \cite{Carlip:2003ne, Lu:2010au}.). Furthermore, it is not
known whether there exists a new globally stable phase in the
present setup other than what have been discussed in the literature so far
\cite{Carlip:2003ne,Lundgren:2006kt,Lu:2010xt}. This was noticed in
\cite{Carlip:2003ne,Lundgren:2006kt} for the charged black hole and
in \cite{Lu:2010xt} for the charged black $p$-brane when $\tilde d
\le 2$.

In this paper, we will resolve the above puzzle and give the new
phase(s) for the $D$-dimensional asymptotically flat stabilized
compact black $p$-brane (for $p>0$) \footnote{For $p = 0$, there is
no bubble phase. The only phases are non-extremal brane and extremal
one, and the phase relation between the two for $p
> 0$ discussed in the text will hold true also for this
case. For simplicity, we focus from now on only for $p
> 0$ case.} by finding two
missing phases, namely, the regular `hot bubble', due to the
existence of ``bubble of nothing" \cite{Witten:1981gj}, and the
extremal brane, each carrying the same flux/charge as the black
$p$-brane. The bubble or the extremal brane \cite{Gibbons:1994ff,
Hawking:1994ii, Hawking:1995fd, Teitelboim:1994az} each can have an
arbitrary period $\beta$ in Euclidean time, analogous to the `hot
flat space' in the chargeless case. As such the black $p$-brane can
make a transition to this bubble or the extremal brane, depending on
whose free energy is smaller, giving an analog of Hawking-Page
transition. As a result, the underlying phase structure is greatly
enriched and many new phase transitions between black
branes, bubbles and extremal brane are revealed, giving a rather
complete phase structure in this setup.

This paper is organized as follows. In section 2, we
will present the basic setup for phase structure of black branes which will be
discussed in the following sections. In section 3, we discuss the
phase structure for the special zero flux/charge case as a warm up
exercise. Here we see how the inclusion of bubble phase will enrich
the previously known phase structure. Section 4 is the main focus of
this paper and we will resolve the aforementioned puzzle for $\tilde
d \le 2$  and find a new global stable (bubble) phase for $\tilde d
> 2$, therefore giving a necessary completion of underlying phase
structure and various new phase transitions including the analog of
Hawking-Page transition among other things. We discuss the results
obtained in this paper and conclude in section 5.

\section{The basic setup}

For the purpose of this paper, let us consider the $D$-dimensional
black $p$-brane metric in Euclidean signature as
\cite{Horowitz:1991cd, Duff:1993ye}, \be\label{blackbrane} ds_{\rm
bl}^2 = \Delta_+ \Delta_-^{-\frac{d}{D-2}}dt^2 +
\Delta_-^{\frac{\tilde
    d}{D-2}}(dx^1)^2 + \Delta_-^{\frac{\tilde d }{D-2}}\sum_{i=2}^p(dx^i)^2
+ \Delta_+^{-1}\Delta_-^{\frac{a^2}{2\tilde d}-1} d\rho^2 + \rho^2
\Delta_-^{\frac{a^2}{2\tilde d}} d\Omega_{\tilde d +1}^2 \ee where
$\Delta_{\pm} = 1 - (r_{\pm}/\rho)^{\tilde d}$, with $r_{\pm}$
($r_+\geq r_-$) related to the mass and the charge of the black
$p$-brane. The horizon occurs at $\rho = r_+$ while the curvature
singularity at $\rho = r_-$. Here $d = 1 + p $, $\tilde d = D - d -
2$ and `$a$' is the dilaton coupling defined by \be
\label{dcoupling} a^2 = 4 - \frac{2d\tilde d}{(D-2)},\ee for
supergravity with maximal supersymmetry. To have a large but finite
Euclidean action for the black brane, the brane directions $x^i$,
with $i=1,\ldots, p$ should be compact (In the metric, $x^1$
coordinate is explicitly isolated for the purpose of constructing
bubble solution later). For the metric \eqn{blackbrane} to be free
from conical singularity at $\rho = r_+$, `$t$' coordinate must be
periodic with periodicity \be \beta^\ast = \frac{4\pi r_+}{\tilde d}
\left(1 - \frac{r_-^{\tilde d}}{r_+^{\tilde d}}\right)^{1/\tilde d -
1/2},\ee the inverse of the black brane temperature at $\rho
=\infty$. With this, the inverse of the local temperature at
 $\rho$ is \be\label{localtemperature}\beta (\rho) = \Delta_+^{1/2}
\Delta_-^{-1/\tilde d} \frac{4\pi \bar r_+}{\tilde d}
\left(1-\frac{\bar r_-^{\tilde d}}{\bar r_+^{\tilde
d}}\right)^{1/\tilde d - 1/2}.\ee However, for the corresponding
extremal brane, `$t$' coordinate can have an arbitrary
period \cite{Gibbons:1994ff, Hawking:1994ii, Teitelboim:1994az} and
this is crucial for the phase transition we will discuss later.
Here the physical radius $\bar \rho \equiv \Delta_-^{a^2/4\tilde
  d}\rho$ can be read from the metric \eqn{blackbrane} and so the physical parameters
$\bar r_{\pm} =
  \Delta_-^{a^2/4\tilde d} r_{\pm}$.
 On the other hand,
the coordinate $x^1$, like other compact coordinates, has arbitrary
local periodicity. By renaming the coordinates $x^1 \to t$ and $t
\to - x^1$ in the black $p$-brane configuration given above, we can
obtain the bubble carrying the same flux/charge in Euclidean
signature with its metric given as\footnote{Though this provides a
simple means to obtain a bubble solution, it can also be obtained by
other means, for example, by solving the underlying equations of
motion. In other words, once the solution is obtained, we can forget
about its connection with the original black brane. Even in the
present context of considering possible allowed phases with the same
boundary data, the parameters $r_+$ and $r_-$, characterizing the
solution, can be different in both the bubble solution \eqn{bubble}
and the black brane \eqn{blackbrane}. As such, the period of $x^1$
(or $t$) in the bubble case is not necessarily the same as that of
$t$ (or $x^1$) in the black brane case. Only when the $r_+$ and
$r_-$ are set to be the same in both cases, the two periods will be
the same which is just a special case. This recognition, as
discussed later in the text, is crucial for the role of the bubble
phase in the phase diagram uncovered in this paper. }
\be\label{bubble} ds_{\rm bb}^2 = \Delta_-^{\frac{\tilde
d}{D-2}}dt^2 + \Delta_+ \Delta_-^{-\frac{d}{D-2}}(dx^1)^2 +
\Delta_-^{\frac{\tilde d }{D-2}} \sum_{i=2}^p(dx^i)^2 +
\Delta_+^{-1}\Delta_-^{\frac{a^2}{2\tilde d}-1} d\rho^2 + \rho^2
\Delta_-^{\frac{a^2}{2\tilde d}} d\Omega_{\tilde d +1}^2, \ee which
is defined only for $\rho \ge r_+$ and is regular if $x^1$ has a
period of \be R^\ast = \frac{4\pi r_+}{\tilde d} \left(1 -
\frac{r_-^{\tilde d}}{r_+^{\tilde d}}\right)^{1/\tilde d - 1/2},\ee
at $\rho = r_+$.
 However, in this case the periodicity of
the coordinate `$t$' as well as the other spatial compact
coordinates remain arbitrary and so it can be in thermal equilibrium
with the cavity in any temperature.

Now in order to study equilibrium thermodynamics
\cite{Gibbons:1976ue} in canonical ensemble, the allowed
configuration (black brane or extremal brane or bubble or
coexistence of two or more of them) should be placed in a cavity
\cite{York:1986it} with fixed physical radius $\bar\rho_B$. The
other fixed quantities are the cavity temperature $1/\beta$, the
physical periodicity of $x^1$, i.e., $R$ (also the physical sizes of
other compact directions), the dilaton value $\bar \phi$ and the
charge/flux enclosed in the cavity $\bar Q_d$. In equilibrium, these
fixed values are set equal to the corresponding ones of the allowed
configuration enclosed in the cavity. For example, we set the charge
\be\label{charge} \bar Q_d = Q_d \equiv \frac{i}{\sqrt{2}
\kappa}\int e^{-a(d)\phi} \ast F_{[p+2]} = \frac{\Omega_{\tilde d
+1}\tilde d}{\sqrt{2}\kappa} e^{-a\bar\phi/2}(\bar r_+ \bar
r_-)^{\tilde d /2},\ee where $\ast$ denotes the Hodge duality and
$F_{[p+2]}= i\tilde d ((\bar r_+ \bar r_-)^{\tilde d}/\bar
\rho^{\tilde d
  +1}) d\bar\rho\wedge dt\wedge dx^1 \wedge \cdots \wedge dx^p$,
the field strength
  for the configuration considered.  In the
charge expression $\Omega_n$ denotes the volume of a unit
$n$-sphere, $\kappa$ is a constant with $1/ (2 \kappa^2)$ appearing
in front of the Hilbert-Einstein action in canonical frame but
containing no asymptotic string coupling $g_s$.
  We have also set
$e^{\bar\phi} = e^{\phi(\bar\rho_B)} \equiv g_s \Delta_-^{a/2}$.  In
canonical ensemble, it is the Helmholtz free energy which determines
the stability of the equilibrium states and is related to the
Euclidean action by $F=I_E/\beta$ in the leading order
approximation. So, in order to understand the phase structure we
will evaluate the action for the black (non-extremal) brane, the
extremal brane and the bubble, with the above
mentioned boundary data. Note that specifying the boundary data does
not necessarily mean that they have the same $\bar r_+$. For
example, in the extremal case, $\bar r_+ = \bar r_-$ are completely
fixed by the given charge.

The Euclidean action for the black $p$-brane has been evaluated in
\cite{Lu:2010xt} by
using standard technique and is given as,
\bea\label{actionbl}
I_E^{\rm bl} &=& \frac{\beta R V_{p-1} \Omega_{\tilde d +1}}{2\kappa^2}\bar
\rho_B^{\tilde d} \left[(2+\tilde
  d)\left(\frac{\Delta_+}{\Delta_-}\right)^{1/2} + \tilde d
  (\Delta_-\Delta_+)^{1/2} - 2(\tilde d +1)\right]\nn
& & -\frac{4\pi R V_{p-1} \Omega_{\tilde d +1}}{2\kappa^2} \bar
r_+^{\tilde d +1} \Delta_-^{-\frac{1}{2} - \frac{1}{\tilde d}}
\left(1 - \frac{\bar r_-^{\tilde
        d}}{\bar r_+^{\tilde d}}\right)^{\frac{1}{2} + \frac{1}{\tilde d}},
\eea with $\Delta_\pm$ taking their respective value at $\bar\rho =
\bar\rho_B$. Since the Helmholtz free energy is given as $F_{\rm bl}
= E_{\rm bl} - TS_{\rm bl}$, so we have $I_E^{\rm bl} = \beta E_{\rm
bl} - S_{\rm bl}$, where $E_{\rm bl}$ is the internal energy and
$S_{\rm bl}$ is the entropy of the black $p$-brane. Thus we identify
the internal energy of the black brane on dividing the first term in
\eqn{actionbl} by $\beta$ and can be checked to match the ADM mass
per unit volume of the black brane \cite{Lu:1993vt} as $\bar \rho_B
\to \infty$. The second term in \eqn{actionbl} is the entropy of the
black brane. Note that we have written the usual compact brane
volume $V_p = R V_{p-1}$. The action for the bubble $I_E^{\rm bb}$
can be obtained simply from \eqn{actionbl} by making the change
$\beta \leftrightarrow R$ and is given as \bea\label{actionbb} I_E^{\rm bb}
&=& \frac{\beta R V_{p-1} \Omega_{\tilde d +1}}{2\kappa^2}\bar
\rho_B^{\tilde d} \left[(2+\tilde
  d)\left(\frac{\Delta_+}{\Delta_-}\right)^{1/2} + \tilde d
  (\Delta_-\Delta_+)^{1/2} - 2(\tilde d +1)\right]\nn
& & -\frac{4\pi \beta V_{p-1} \Omega_{\tilde d +1}}{2\kappa^2} \bar
r_+^{\tilde d +1} \Delta_-^{-\frac{1}{2} - \frac{1}{\tilde d}}
\left(1 - \frac{\bar r_-^{\tilde
        d}}{\bar r_+^{\tilde d}}\right)^{\frac{1}{2} + \frac{1}{\tilde
        d}}.
\eea As for the bubble since there is no entropy, we have $I_E^{\rm
bb} = \beta E_{\rm bb}$ and so we can identify the internal energy
of the bubble on dividing its action by $\beta$ and again can be
checked to have the correct ADM mass per unit volume of the bubble
as $\bar \rho_B \to \infty$. However, for the black brane in
canonical ensemble the local temperature $1/\beta(\bar \rho_B)$ is
determined by $r_+, Q_d, \bar \rho_B$ as given before  and the
periodicity of $x^1$ is arbitrary whereas for the bubble the local
periodicity of $x^1$, i.e. $R(\bar \rho_B)$, is similarly determined
by the corresponding quantities and the temperature is now
arbitrary. On-shell, they are all set equal to the corresponding
fixed boundary values.

Now for convenience,  we will work instead with the reduced action
defined as $ \tilde I_E^{\rm {bl,\,bb}} (z) = 2\kappa^2
I_E^{\rm{bl,\, bb}}/((4\pi)^2 \bar \rho_B^{\tilde d +2} V_{p-1}
\Omega_{\tilde d +1}) = G_q(z)$ with $G_q(z)$ defined as,
\bea\label{actionreduced1} G_q(z) &=& - \bar b \bar R \left[(\tilde
d + 2) \left(\frac{1 - z}{1 - \frac{q^2}{z}}\right)^{1/2} + \tilde d
(1 - z)^{1/2}\left( 1 - \frac{q^2}{z}\right)^{1/2} - 2 (\tilde d +
1)\right]\nn &\,& - \bar l\,  z^{1 + 1/\tilde d} \left(\frac{1 -
\frac{q^2}{z^2}}{1 - \frac{q^2}{z}}\right)^{1/2 + 1/\tilde d}, \eea
where we have defined $z=(\bar r_+/\bar \rho_B)^{\tilde d} < 1$ and
$z=x$ for the black brane and $z=y$ for the bubble since the two
need not have the same $\bar r_+$. Further we have defined $\bar b =
\beta/(4\pi \bar \rho_B)$, $\bar R = R/(4\pi \bar \rho_B)$ and $\bar
l = \bar R$ for the black brane and $\bar l = \bar b$ for the
bubble. Also, $q = (Q_d^*/\bar \rho_B)^{\tilde d}$, where $Q_d^* =
[(\sqrt{2}\kappa \bar Q_d)/(\Omega_{\tilde d +1} \tilde
d)]^{1/\tilde d}$. In terms of these new fixed parameters we have,
  \be \Delta _+ = 1-z,\qquad \Delta_-
= 1-q^2/z,\qquad 1- \frac{\bar r_-^{\tilde d}}{\bar r_+^{\tilde d}}
= 1 - q^2/z^2,\ee and they have been used in \eqn{actionreduced1}.
In writing $\Delta_-$ we have used the fact that since $\bar Q_d$ is
fixed, $\bar r_-$ is not an independent parameter but can be
expressed in terms of $\bar r_+$. Given $(Q_d^*)^2/\bar r_+^2 = \bar
r_-/\bar r_+ \le 1$,  so $z \ge q$. For extremal branes, $\bar r_- =
\bar r_+$ and so $z = q$. The reduced action for extremal
  branes is now \be\label{extremal} \tilde I^{\rm extremal}_E
= \bar b \bar R \tilde d q,\ee
   determined completely by the boundary data. However,
  for non-extremal brane or bubble, we have
 a variable $z$ lying between $q < z < 1$. Now $\left.d G_q(z)/d z
\right|_{z = \bar z} =0$ gives the equation of state
 $\bar m = m_q(\bar z)$, where, \be\label{onshell} m_q(z) =
\frac{1}{\tilde d} \frac{z^{1/\tilde d}(1-z)^{1/2}}{\left(1 -
    \frac{q^2}{z^2} \right)^{\frac{1}{2}-\frac{1}{\tilde d}} \left(1 -
    \frac{q^2}{z}\right)^{1/\tilde d}}.
\ee If \be \label{extremalcondition} \left. d^2 G_q (z)/d
z^2\right|_{z = \bar z} \propto - \left. d m_q (z)/d z\right|_{z =
\bar z}
> 0,\ee  at $z = \bar z$, where $\bar z$ is determined from $\bar m
= m_q (\bar z)$, we get a local minimum of free energy. So the
negative slope of $m_q (z)$ determines the local stability of the
underlying system. In the above $\bar m = \bar b, \, m_q(z = x) =
b_q(x)$ for the black brane and $\bar m = \bar R, \, m_q(z = y) =
R_q(y)$ for the bubble (Note that we always have $\bar m \bar l =
\bar b \bar R$.).

The phase structure of $G_q(z)$ corresponding to the black $p$-brane
has been analyzed in \cite{Lu:2010xt}. The bubble also has exactly
similar phase structure, however, the relevant quantities here are
 $\bar R$ and $R_q (y)$ instead. For now, we just need to compare
the free energies among the black brane,
the bubble, at their respective global minimum, and the extremal
brane, all with the same boundary data.  The analog of Hawking-Page
transition in either case and the final stable state will be
determined by  the smallest free energy of these phases. For this
purpose we need the respective on-shell reduced free energy
explicitly. It is \be \label{reducedfreeenergy} \tilde F^{\rm bl,\,
bb} (\bar z) \equiv \frac{\tilde I_E^{\rm {bl,\, bb}} (\bar z)}{\bar
b} = - \bar R F_q(\bar z),\ee with \be \label{fq} F_q( z) = 2
\left(\frac{1 - z}{1 - \frac{q^2}{z}}\right)^{1/2} + \tilde d
\left(\frac{1 - \frac{q^2}{z}}{1 - z}\right)^{1/2} + \tilde d
\left(1 - z\right)^{1/2} \left(1 - \frac{q^2}{z}\right)^{1/2}
 - 2 (\tilde d + 1).\ee In the above, we have used the on-shell
condition $\bar
 m = m_q (\bar z)$ with $\bar z$ lying between $q < \bar z < 1$.
 Given the boundary data $\bar b$
and $\bar R$, the on-shell free energy of non-extremal brane is
$\tilde F^{\rm bl}
(\bar x) = - \bar R F_q (\bar x)$, with $\bar x$ determined by $\bar
b = b_q (\bar x)$, while the on-shell free energy of bubble is $\tilde
F^{\rm bb} (\bar y) = - \bar R F_q (\bar y)$, but now with $\bar y$
determined by $\bar R = R_q (\bar y)$. So $\tilde F^{\rm bl} (\bar x)$ and
$\tilde F^{\rm bb} (\bar y)$ actually have different dependence on
their respective on-shell variable even though both have the same
functional form $- \bar R F_q (\bar z)$ in appearance. So
their profiles are in general different. For example,
 $\tilde F^{\rm bl} (\bar x = q) = \bar R \tilde d q$ (giving the extremal
brane free energy) and
$\tilde F^{\rm bl} (\bar x \to 1) \to - \infty$ while $\tilde F^{\rm
bb} (\bar y \to q) \to \infty$ and $\tilde F^{\rm bb} (\bar y \to 1)
\to - 1$.  Note that at the two ends $F_q(q) = -\tilde d q$
 and $F_q (1) =
\infty$. In spite of their difference, both $\tilde F^{\rm bl}
(\bar z)$ and $\tilde F^{bb} (\bar z)$ have the same characteristic
behavior as that of $m_q (\bar z)$, and this can be understood from
the following
relation (note its difference from \eqn{extremalcondition}),\be \label{freeemc} \frac{d \tilde F^{\rm bl, \,
bb} (\bar z)} {d \bar z} \propto \frac{d m_q(\bar z)}{d \bar z}, \ee
 where we
have used  \eqref{onshell}, \eqref{reducedfreeenergy}
and the above $F_q (z)$ (In the case of bubble, we also need to
consider the contribution from $d \bar R/d\bar y = d R_q (\bar
y)/d\bar y$). For
example, the maximum or minimum (if exists at all) for each of these
functions occurs at the same $z_{\rm max}$ or $z_{\rm min}$.
  The above enables us, in the present context, to
make use of the extrema (if any) and behavior of $m_q(z)$ studied in
\cite{Lu:2010xt} to compare the free energies at the respective
global minimum.

We are interested in the region for which the corresponding
free energy will be at least locally stable, i.e.,
$d m_q (z)/d z|_{z = \bar z} < 0$.
Equation  \eqref{freeemc} immediately tells us that the on-shell
$\tilde F^{\rm bl} (\bar z)$ and  $\tilde F^{\rm bb} (\bar z)$ along
with $m_q (\bar z)$ will decrease in this region. For $ m_q
(\bar z)$, $\bar b = b_q (\bar x)$ and $\bar R = R_q (\bar y)$ then
imply that in each of the cases $\bar b
> \bar R, \, \bar b = \bar R$ and $\bar b < \bar R$ we will get
$\bar x < \bar y, \, \bar x = \bar y$ and $\bar x > \bar y$,
respectively. Further, we can use the property of either
$\tilde F^{\rm bl} (\bar x)$ or
$\tilde F^{\rm bb} (\bar y)$ as a decreasing function to determine
which of the phases -- the brane or the bubble have smaller free
energy for given $\bar b$
and $\bar R$. Let us take a particular case for illustration, say
$\bar b
> \bar R$, implying $\bar x < \bar y$. So we have $\tilde F^{\rm bl}
(\bar x) > \tilde F^{\rm bl} (\bar y) = \tilde F^{\rm bb} (\bar y)$,
which says now that the bubble has smaller free energy. Alternatively,
this can
also be determined using $\tilde F^{\rm bb}$ as follows:
$\tilde F^{\rm bb} (\bar y) < \tilde F^{\rm bb} (\bar x) = - \bar b
F_q (\bar x) < - \bar R F_q (\bar x) = \tilde F^{\rm bl} (\bar x)$.
In the following two sections, we will use $\tilde F^{\rm bl} (\bar
z)$ as a decreasing function in the region of interest to determine the
global stable phase, which
appears slightly more straightforward than using $\tilde F^{\rm bb}
(\bar z)$ instead.

With the above preparation, we are now ready to discuss the phase
structure for various cases as promised in the Introduction.

\section{The zero flux/charge case}

Let us discuss the zero charge/flux case first. As discussed in
\cite{Lu:2010xt}, the function $m_0( z)=0$ at the two ends $
z=0,\,1$ (look at eq.\eqref{onshell}) and has a maximum
\be m_{\rm max} = \frac{1}{\sqrt{2\tilde
d}}\left(\frac{2}{\tilde d +2}\right)^{\frac{1}{2} + \frac{1}{\tilde
d}}\ee in between at $z_{\rm max} = 2/(\tilde d +2)$ (see the second
graph in Fig. 1).
\begin{figure}
\psfrag{A}{$\tilde F^{\rm bl} (\bar z)$} \psfrag{B}{$0$}
\psfrag{C}{$z_{\rm max}$} \psfrag{D}{$z_g$} \psfrag{E}{$1$}
\psfrag{F}{$\bar z$}\psfrag{A1}{$m_0 (z)$}\psfrag{B1}{$m_{\rm
max}$}\psfrag{B2}{$\bar m$}\psfrag{B3}{$m_g$}\psfrag{n}{$z_{\rm
max}$}\psfrag{l}{$\bar
z_2$}\psfrag{k}{$z_g$}\psfrag{F1}{$z$}\psfrag{m}{$\bar z_1$}
\begin{center}
  \includegraphics{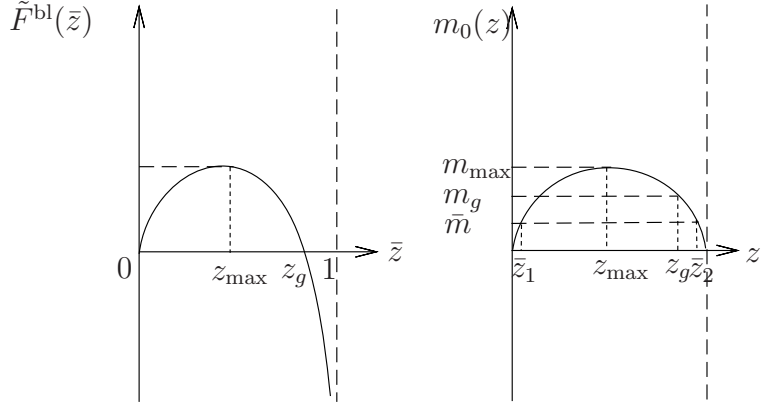}
  \end{center}
  \caption{ $\tilde F^{\rm bl} (\bar z)$ vs $\bar z$
and $m_0 (z)$ vs $z$ for zero flux/charge case.}
\end{figure}
So, above $m_{\rm max}$ there is no black brane or bubble phase and
the system will be in `hot flat space' phase (with zero free
energy). But below $ m_{\rm max}$, the $\bar m = m_0 (\bar z)$
always gives a local minimum of free energy at each $\bar z_2$ lying
in $z_{\rm max} < \bar z_2 < 1$ since there $ d m_0 (z)/d z|_{z =
\bar z_2} < 0$, implying a locally stable black brane or bubble. The
locally stable configuration becomes globally stable when \be \bar m
< m_g = \frac{1}{\tilde d +2} \left(\frac{4(\tilde d +1)}{(\tilde d
+2)^2}\right)^{1/\tilde d},\ee (i.e, $b_g = R_g = m_g$) at $\bar
z_2$ lying in $1> \bar z_2
> z_g $ with
\be z_g = \frac{4(\tilde d +1)}{(\tilde d +2)^2},\ee since its free
energy is now negative. Here $z_g$ is determined by setting the free
energy to zero at $\bar z_2 = z_g > z_{\rm max}$. In other words,
the free energy has its positive maximum value at $z_{\rm max}$,
decreases from there to zero at $z_g (> z_{\rm max})$, then
continues to decrease to negative and finally reaches $- \infty$ at
$\bar z_2 =1$ (see the first graph for black brane in Fig. 1).
So from $\bar z_2 =
z_{\rm max} =2/(\tilde d +2)$ to $\bar z_2 =1$, the free energy is a
monotonically decreasing function and so, greater $\bar z_2$ (i.e.,
$\bar x_2$ or $\bar y_2$) will give smaller free energy for the
black brane or the bubble. Keeping this in mind we consider the
following cases: 1) If $\bar R$, $\bar b > b_g = R_g$, the black
brane or bubble phase is at most locally stable with a positive free
energy and the locally stable phase will make a Hawking-Page phase
transition to the globally stable `hot flat space' phase. If $\bar R
> \bar b = b_g$ (or $\bar b > \bar R = R_g$), the bubble
(or the black brane) is at most locally stable  and the locally
stable phase will make also a Hawking-Page transition to the
globally stable phase which is now a coexisting phase of  both the
black brane (or the bubble) and the `hot flat space'. 2) If $\bar R
= \bar b = b_g = R_g$, the three phases -- the bubble, the black
brane and the `hot flat space' coexist and the transition between
any two of them is a first order one since the first derivative of
free energy has a discontinuity when the transition occurs. 3) If
$\bar R < R_g$ and $\bar R < \bar b$, the bubble is the globally
stable phase. If $\bar b < b_g $ and $\bar b < \bar R$, black brane
is the globally stable phase instead. If $\bar b = \bar R < b_g =
R_g$, the bubble and the black brane coexist and are globally
stable.

\section{The non-zero flux/charge case}

We now move on to the case when the charge enclosed in the cavity is
fixed but non-zero. As shown in \cite{Lu:2010xt}, when the charge is
non-zero, there exists a critical charge $q_c$ at which the first
and second derivatives of $m_q (z)$ with respect to $z$ at $z = z_c$
vanish and these will determine completely $q_c$, $z_c$ and $m_c$.
The underlying phase structure crucially depends on whether $q
> q_c$, $q = q_c$ and $q < q_c$, so we discuss them in turns in the
following for $\tilde d
> 2$ (The $\tilde d \le 2$ cases are different and will be discussed
afterwards.): i) For $q > q_c$, there is no extrema for either $m_q(
z)$  or free energy $\tilde F^{\rm bl,\, bb} (\bar z)$, with each
decreasing monotonically in the region of $q < z,\, \bar z < 1$ (see
the first graph in Fig. 2 for $m_q (z)$ and in Fig. 3
for $\tilde F^{\rm bl} (\bar z)$).
\begin{figure}
\psfrag{A}{$m_q (z)$} \psfrag{A0}{$m_q (z)$} \psfrag{B}{$z$}
\psfrag{C}{$\bar m$} \psfrag{C0}{$\bar m$} \psfrag{D}{$\bar z$}
\psfrag{E}{$m_{\rm max}$} \psfrag{F}{$m_{\rm min}$} \psfrag{A1}{$q
>q_c$}\psfrag{A2}{$q<q_c$}\psfrag{D1}{$q$}\psfrag{D2}{$1$}\psfrag{C1}{$\bar
z_1$}\psfrag{C2}{$\bar z_2$}\psfrag{C3}{$\bar
z_3$}\psfrag{F1}{$z_{\rm min}$}\psfrag{F2}{$z_{\rm max}$}
\begin{center}
  \includegraphics{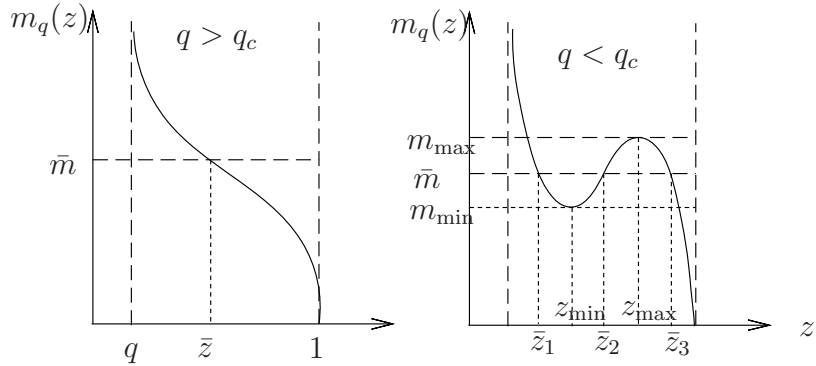}
  \end{center}
  \caption{The typical behavior of $m_q (z)$ vs $z$ for
$q > q_c$ and $q < q_c$}
\end{figure}
The slope of $m_q ( z)$ is always negative and therefore the
on-shell $\tilde F^{\rm bl, \, bb} (\bar z)$ is a local minimum. So,
we have the following cases to consider. If $\bar b
> \bar R$, the respective $\bar m = m_q (\bar z)$ implies
$\bar x < \bar y$ and so,
$\tilde F^{\rm bl} (\bar x) > \tilde F^{\rm bb} (\bar y)$,
 i.e., bubble is the globally stable phase.
In other words, an analog of Hawking-Page transition will take the
locally stable black brane phase to the globally bubble phase.
Similarly, if $\bar b < \bar R$, the black brane is the globally
stable phase. But if $\bar b = \bar R$, the bubble and the black
brane phases coexist and the transition between the two is a first
order one due to the change of entropy involved. ii) For $q=q_c$, we
have $\bar b_c = \bar R_c$ and $x_c = y_c$, and the two phases are
both globally stable and can coexist at the critical point and the
transition between the two is also a first order one. iii) For $q <
q_c$, the situation is a bit involved. As noted in \cite{Lu:2010xt},
in this case, the function $m_q(z)$ does not decrease monotonically
in the region $q < z < 1$, but in between there is a minimum $m_{\rm
min}$ at $z = z_{\rm min}$ and a maximum $m_{\rm max}$ at $z =
z_{\rm max}$ (see the second graph in Fig. 2). Since $m_q(z)$ starts
at infinity at $z=q$ and goes to zero at $z=1$ (see
eq.\eqn{onshell}), so $z_{\rm min} < z_{\rm max}$. Then as shown in
\cite{Lu:2010xt}, in the range $q < \bar z < z_{\rm min}$, and
$z_{\rm max} < \bar z < 1$ the black branes or the bubbles are
locally stable since each $\bar z$ gives a local minimum of free
energy, whereas in $z_{\rm min} < \bar z < z_{\rm max}$ they are
unstable since each $\bar z$ gives a maximum of free energy.
So, for $m_{\rm min} < \bar m < m_{\rm max}$, $\bar m = m_q (\bar
z)$ gives three black brane or bubble phases with three solutions,
say, $\bar z_1 < \bar z_2 < \bar z_3$, where $\bar z_1$ (small black
brane or bubble) and $\bar z_3$ (large black brane or bubble)
correspond to the locally stable phases and $\bar z_2$ corresponds
to unstable phase (see the second graph in Fig. 2). Among
the locally stable phases the system would prefer to be in the phase
of lowest free energy. Now as demonstrated in \cite{Lu:2010xt}, in
between $m_{\rm max}$ and $m_{\rm min}$ and for given $q$, there
exists a $\bar m$ denoted by $m_t (q, \tilde d)$ (here $b_t = R_t$),
a function of $\tilde d$ and charge $q$ only, where the large black
brane (bubble) of size $x_{3t}$ ($y_{3t}$) has the same free energy
($\tilde F^{\rm bl, \, bb} (z_{3t}) = \tilde F^{\rm bl,\, bb}
(z_{1t})$) as the small black brane (bubble) of size $x_{1t}$
($y_{1t}$) and so they coexist. Above $m_t$ and below $m_{\rm max}$
the small black brane (bubble) is globally stable and below $m_t$
and above $m_{\rm min}$, the large black brane (bubble) is the
globally stable phase if the black brane or the bubble is assumed to
be the only phase. Given what has been said about either black
branes or bubbles, we now determine the globally stable phase with
the lowest free energy between black branes and bubbles with the
same boundary data. Given that $\tilde F^{\rm bl, \, bb} (\bar z)$
decreases monotonically in the range $q < \bar z < z_{\rm min}$ and
$z_{\rm max} < \bar z < 1$ (see the second graph in Fig. 3), so when
$\bar b, \bar R
> b_t = R_t$ or $\bar b , \bar R < b_t = R_t$, which phase is
globally stable can be discussed following $q > q_c$ case and will
not be repeated here. When $\bar b = \bar R = b_t = R_t$, all four
phases i.e., the small black brane, the large black brane, the small
bubble and the large bubble coexist. The transition from one phase
to the other is a first order one (when it is between small and
large black branes or between small and large bubbles) ending at a
second order critical point when $q = q_c$ and a first order one
(when it is between a black brane and a bubble) even at $q = q_c$
(due to a change of entropy there). The reason that this transition
is always a first order one when $q < q_c$ is again due to the
discontinuity of the first derivative of free energy when the
transition occurs. When $\bar b
> b_t = R_t$ and $\bar R < b_t = R_t$, $\bar x < x_{1t} < x_{3t} = y_{3t} <
\bar y$ and so, $ \tilde F^{\rm bl} (\bar x) > \tilde F^{\rm bl}
(x_{1t}) = \tilde F^{\rm bl} (x_{3t}) = \tilde F^{\rm bb} (y_{3t}) >
\tilde F^{\rm bb} (\bar y)$, therefore, the large bubble is the
globally stable phase. Similarly, when $\bar b < b_t = R_t$ and
$\bar R > b_t = R_t$, the large black brane is the globally stable
phase.
\begin{figure}
\psfrag{A}{$\bar z$} \psfrag{B}{$\tilde F^{\rm bl} (\bar z)$}
\psfrag{C} {$z_{t1}$} \psfrag{D}{$z_{\rm min}$} \psfrag{E}{$z_{\rm
max}$} \psfrag{F}{$z_{t3}$} \psfrag{G}{($ q > q_c$)} \psfrag{H}{($q
< q_c$)} \psfrag{A1}{$q$} \psfrag{A2}{$1$}
\begin{center}
  \includegraphics{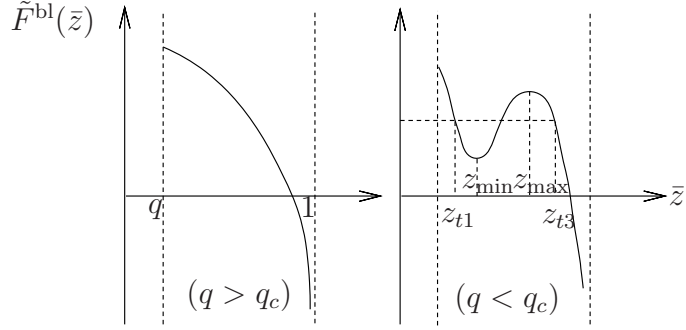}
  \end{center}
  \caption{The typical behavior of reduced free energy
$\tilde F^{\rm bl} (\bar z)$ vs $\bar z$ for $q > q_c$
and $q < q_c$.}
\end{figure}

The extremal brane phase with the same boundary data has always the
largest free energy in any of the above three cases and therefore
can never be a globally stable phase.

We now move on to the $\tilde d \le 2$ cases which have
similarities with the zero charge case. Let us consider $\tilde d
=1$ case first. The phase analysis for this case can be discussed
following the chargeless case if we replace the `hot flat space'
there by the present extremal brane phase and $1 > z > 0$ for $m_0
(z)$ there by $1 > z > q$ for $m_q (z)$ here and will not be
repeated here. The present $m_{\rm max}$ and $z_{\rm max}$ can be
determined now from $m_q (z)$ similarly but they cannot be given
analytically as demonstrated in \cite{Lu:2010xt}. Instead, we will
give their numerical values for a few selected values of
$q$ along with the values of $m_g$ and $z_g$ in the following table.
The $z_g$ falling in the range $1
> z_g
> z_{\rm max}$ can be determined from the following equation,
\be\label{tilded1} 4 - q = 2 \left(\frac{1 - \bar z}{1 -
\frac{q^2}{\bar z}}\right)^{1/2} + \left(\frac{1 - \frac{q^2}{\bar
z}}{1 - \bar z}\right)^{1/2} +  (1 - \bar z)^{1/2} \left(1 -
\frac{q^2}{\bar z}\right)^{1/2},\ee  which is obtained by equating
the system free energy at $\bar z = z_g$ with the corresponding
extremal brane free energy at $\bar z = q$, i.e., $ F_q (\bar z = z_g) =
F_q(\bar z=q)=q$. This is actually a quartic equation but one can
check, as expected, that $\bar z = q$ is a solution. So this
equation can be reduced to a third-order equation as \be 9 \bar z^3 - (5
q^2 - q + 8)\bar z^2 + (3 q + 4) q^2 \bar z - 4 q^3 = 0,\ee which
can be solved analytically and has only one real positive solution,
giving $z_g$. The solution for $z_g$ is complicated and not very
illuminating and for this
reason we will not give its explicit analytic expression here. As
mentioned above, we will list instead its value  for each
  given $q$ along with the corresponding $m_g$ for
  $q$ starting at $q = 0.1$ with an increment of $0.1$ up to $q = 0.9$
  in the table below.
  \vspace{.2cm}

\begin{center}
\begin{tabular}{|c|c|c|c|c|}
\hline\hline
$q$ & $z_g$ & $m_g$ & $z_{\rm max}$ & $m_{\rm max}$ \\
\hline\hline
0.1 & 0.878471 & 0.307757 & 0.668371& 0.386344\\
\hline
0.2 & 0.870088 & 0.319918 & 0.673943 &0.390683\\
\hline
0.3 & 0.864945 &0.332758  &0.673943& 0.397903\\
\hline
0.4 & 0.864458 & 0.346219 &0.684781& 0.407890\\
\hline
0.5 & 0.870015 &0.360200  &0.731065&0.420323\\
\hline
0.6&0.882600  &0.374565&0.769645&0.434666\\
\hline
0.7&0.902505&0.389161&0.817842&0.450290\\
\hline
0.8 &0.929316&0.403845& 0.873604&0.466636\\
\hline
0.9 & 0.962160&0.418491&0.934861&0.483295\\
\hline\hline
\end{tabular}
\end{center}
\vspace{.2cm} From the above table, one can see clearly that $1 >
z_g > z_{\rm max}$ and $m_g < m_{\rm max}$ as expected.

 Let us now discuss
$\tilde d =2$ case. For this case, as shown also in
\cite{Lu:2010xt}, there exists not only (like $\tilde d
> 2$ case) a critical charge with\footnote{
  However, now $z_c = q_c$, occurring at the small end, and therefore,
unlike $\tilde d > 2$, there is now no critical behavior since we don't have
  a stable small brane (or bubble) phase.} $q_c = 1/3$, but also (like
$\tilde d =1$ case) a maximum of $m_q (z)$ at $z_{\rm max}$. For $q
\geq q_c$, we have $m_{\rm max} = m_q (q) = \sqrt{q}/2$ at $z_{\rm
max} = q$ (see the first graph in Fig. 4).
\begin{figure}
\psfrag{A}{$m_q (z)$} \psfrag{B}{$z$}\psfrag{C}{$\bar
m$}\psfrag{C0}{$m_g$}\psfrag{C1}{$m_{\rm max}$}\psfrag{D}{$\bar
z$}\psfrag{D1}{$q$}\psfrag{D2}{$1$}\psfrag{E1}{$z_{\rm
max}$}\psfrag{E2}{$z_g$}\psfrag{A1}{$q \ge q_c$}\psfrag{A2}{$q <
q_c$}
\begin{center}
  \includegraphics{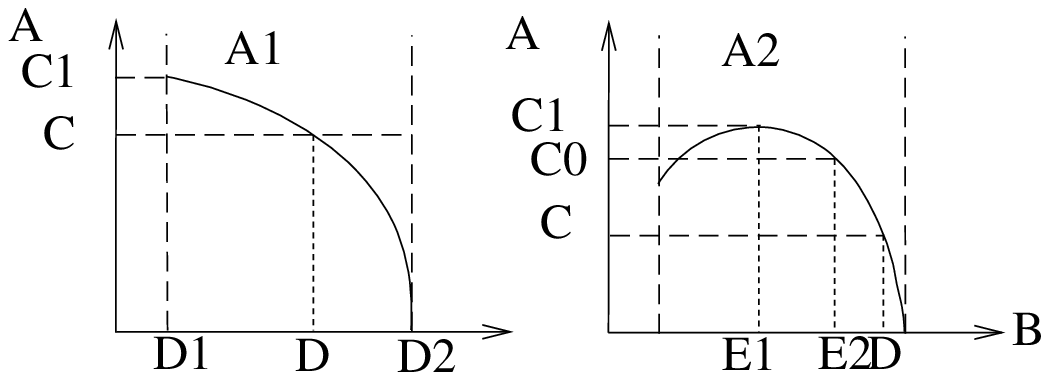}
  \end{center}
  \caption{The typical behavior of $m_q (z)$ vs $\bar z$ for $q > q_c$
and $q < q_c$ when $\tilde d = 2$.}
\end{figure}
When $\bar m > m_{\rm max}$, the extremal brane is the global phase.
But below $m_{\rm max}$, this case is similar to the analysis of
$\tilde d
> 2$ and will not be repeated here. When $q < q_c = 1/3$, we have now
\bea  m_{\rm max}  &=& \frac{\left(1 + 3 q^2 +
 \sqrt{(1 - q^2)(1 - 9 q^2)}\right)^{1/2} \left(3\sqrt{1 - q^2} - \sqrt{1 - 9
 q^2}\right)}{8 \sqrt{2}},\nn
  z_{\rm max} &=& \frac{1 + 3 q^2 + \sqrt{(1 - q^2) (1 - 9
q^2)}}{4},\eea with $m_{\rm max} > m_q (q) = \sqrt{q}/2$ and $z_{\rm
max} > q$ (see the second graph in Fig. 4). This subcase is however
similar to the discussion of $\tilde d =1$. Again when $\bar m >
m_{\rm max}$, the extremal brane is the stable phase. When $\bar b <
b_{\rm max}$ and $\bar R > R_{\rm max}$ (or $\bar b > b_{\rm max}$
and $\bar R < R_{\rm max}$) there exists a locally stable black
brane (or bubble). As in $\tilde d =1$ case, we now need to
determine $x_g$ (or $y_g$) below which locally stable black brane
(or bubble) becomes globally stable. Exactly as in $\tilde d = 1$
case, $z_g$ is now determined by
 the following equation, \be \label{tilded2} 3 - q = \left(\frac{1 -
\bar z}{1 - \frac{q^2}{\bar z}}\right)^{1/2} + \left(\frac{1 -
\frac{q^2}{\bar z}}{1 - \bar z}\right)^{1/2} +  (1 - \bar z)^{1/2}
\left(1 - \frac{q^2}{\bar z}\right)^{1/2}. \ee This equation, unlike
in the previous case, can be solved exactly and has apart from the
trivial solution at $\bar z=q$, two other real solutions. Only one
of them lies in the range $z_{\rm max} < z_g < 1$ and is given as
\be\label{zg} z_g = \frac{3 - 2 q + 3 q^2 + (1 + q)\sqrt{3 (3 - q)(1
- 3q)}}{8},\ee with $1> z_g > z_{\rm max}$  With this $z_g$, we have
\be\label{bgrg} m_g = \frac{\left[5 - 3 q - \sqrt{3(3 - q)(1 -
3q)}\right]\left[3 - 2 q + 3 q^2 + (1 + q)\sqrt{3(3 - q)(1 - 3
q)}\right]^{1/2}}{16 \sqrt{2}}.\ee Now the discussion for globally
stable phase and the analog of Hawking-Page transition are identical
to the chargeless case if we replace the `hot flat space' there by
the present extremal brane phase and therefore will not be repeated.

\section{Discussion and conclusion}
In this paper, we find that a necessary completion of phase
structure of charged black $p$-branes in canonical ensemble requires
two additional phases, namely, the bubble and the extremal brane.
This finding solves the puzzle about the missing phase(s) for
$\tilde d \le 2$ and gives a new phase for $\tilde d > 2$. We also
find an analog of Hawking-Page transition among many new phase
transitions revealed, giving now a much enriched and complete
underlying phase structure (assuming no other new phases present)
which can have a coexistence of up to four individual phases. The
phase transitions are always the first-order ones. We also find that
the extremal brane cannot be a stable phase for $\tilde d > 2$ but
is vital to the completion of phase diagram for $\tilde d \le 2$.
These results are obtained solely on the basis of the free energy
criterion which is used to justify the underlying stability of the
various phases.

Before closing this section, we need to clarify which spin structure
the extremal brane should take so that no inconsistency arises. For
this, let us be somewhat specific in the present context. We focus
on the relevant $(t, x^1, \rho)$ directions. For the non-extremal
branes, the topology is $R^2 \times S^1$ with the $S^1$ denoting the
$x^1$-circle while for the bubble it is $S^1 \times R^2$ with now
$S^1$ denoting the $t$-circle. Though the two have the same topology
in Euclidean signature, they are quite different in Lorentzian
signature since their spacetime structures are different. Also the
former has a non-vanishing entropy while the latter has zero
entropy. In spite of their differences, the two share one common
feature that there is only one spin-structure allowed corresponding
to the fermions being anti-periodic along their respective $S^1$.
However, for the extremal brane, the topology is $S^1 \times S^1
\times R$ and in general there are two spin structures, one for
fermions being periodic and the other for fermions being
anti-periodic, along either circle. The extremal $p$-branes, when
considered in isolation,  are the 1/2-BPS objects in string/M
theory, which preserve one half of the spacetime supersymmetries. So
the natural choice for the spin structure is with periodic fermions.
If this were the case in the present context, one would have
difficulty in understanding\footnote{No such issues would arise for
$\tilde d > 2$ since there is no phase transition between the
extremal brane and the non-extremal brane (or the bubble).} the
phase transitions between the extremal brane and the non-extremal
brane (or bubble) for $\tilde d \leq 2$, vital for the completion of
phase diagram, since the two have different spin structure as
discussed above. However, the present setup of placing the extremal
brane in a cavity with a fixed finite temperature in canonical
ensemble breaks all the underlying supersymmetries and therefore
requires the spin structure with anti-periodic fermions
\cite{Horowitz:2005vp} for the extremal brane\footnote{ The
discrepancy in the entropy between the semi-classical calculation
\cite{Gibbons:1994ff, Hawking:1994ii, Teitelboim:1994az} and the
microstate counting in string theory has been addressed recently in
\cite{Carroll:2009maa}. It was pointed out there that the
non-vanishing of the entropy is due to a different reason and the
semi-classical result of vanishing entropy for the extremal black
hole may still hold. The associated issue as discussed in
\cite{Horowitz:2005vp} is that the tachyon condensation may modify
the semi-classical topology for the extremal brane when the circle
size is of the order of string scale for which the
curvature can be large and the semi-classical analysis may break
down. While these issues appear to be not completely settled
at present, however, we may
re-interpret the modification of the topology of the time circle (or the
spatial circle) due to the tachyon condensation, as the underlying
dynamical process of the phase transition from the extremal
brane to the non-extremal brane (or the bubble) (for $\tilde d \le 2$)
if the free energy indeed favors this process. Though difficult to
imagine, the reverse process (if the free energy favors) may give
the transition from the non-extremal brane (or bubble) to the
extremal brane.} instead. So spin structure wise, there is no
paradox arising and the free energy criterion is as usual a suitable
means to obtain the phase structure and phase transitions as given
in this paper.

\section*{Acknowledgements:}

JXL acknowledges support by grants from the Chinese Academy of
Sciences, a grant from 973 Program with grant No: 2007CB815401 and a
grant from the NSF of China with Grant No : 10975129.

\end{document}